\documentclass{article}

\usepackage{verbatim}
\usepackage{amsmath,amsfonts}
\usepackage{hyperref}
\usepackage{a4wide}
\usepackage{graphicx}
\usepackage[sort&compress,numbers,square]{natbib}

\usepackage{color}
\definecolor{mygrey}{rgb}{0.8,0.8,0.8}
\definecolor{mydarkdarkred}{rgb}{0.1,0,0}
\definecolor{mydarkdarkblue}{rgb}{0,0,0.1}
\newcommand{\myurl}{\mbox{\url{http://web.physik.rwth-aachen.de/download/valkenburg/ColLambda/}}}

\newcommand{\Mtsq}{\tilde M^2}
\newcommand{\sqrtMtsq}{\tilde M}
\newcommand{\Mtsqsq}{\tilde M^4}

\begin{document}
\title{Complete solutions to the metric of spherically collapsing dust in an expanding spacetime with a cosmological constant}
\date{7 July 2012}
\author{Wessel Valkenburg\footnote{valkenburg@lorentz.leidenuniv.nl} \\Instituut-Lorentz for Theoretical Physics, Universiteit Leiden \\ Niels Bohrweg 2, 2333 CA Leiden, The Netherlands\\
Institut f\"ur Theoretische Physik, Universit\"at Heidelberg \\ Philosophenweg 16, D-69120 Heidelberg, Germany}

\maketitle

\begin{abstract}
We present elliptic solutions to the background equations describing the Lema\^itre-Tolman-Bondi (LTB) metric as well as the homogeneous Friedmann equation, in the presence of dust, curvature and a cosmological constant $\Lambda$. 
For none of the presented solutions any numerical {\em integration} has to be performed.
All presented solutions are given for expanding and collapsing phases, preserving continuity in time and radius; both radial and angular scale functions are given. 
Hence, these solutions describe the complete spacetime of a collapsing spherical object in an expanding universe, as well as those of ever expanding objects.
 In the appendix we present for completeness a solution of the Friedmann equation in the additional presence of radiation, only valid for the Robertson-Walker metric.
\end{abstract}

\section{Introduction}
Cosmic structures today have entered the non-linear regime. They can not on all scales be described by a linear perturbation theory on top of the Friedmann-Lema\^itre-Robertson-Walker (FLRW) metric~\cite{springerlink:10.1007/BF01332580,Friedmann:1924bb,1931MNRAS..91..483L,1935ApJ....82..284R,Walker01011937}. The simplest step beyond linear perturbation theory is to look at separate patches, describing the evolution in each patch as a closed system, insensitive to other perturbations outside the patch. Over-densities can be studied in the approximation of spherical collapse, where underdensities expand and potentially become spherical
 voids. Either way, these spherically symmetric configurations, whether matching to a surrounding FLRW metric or not, are described by the Lema\^itre-Tolman-Bondi (LTB) metric~\cite{1933ASSB...53...51L,Tolman:1934za,Bondi:1947av},
 \begin{align}
ds^2=-dt^2+S^2(r,t)dr^2+R^2(r,t)(d\theta^2+\sin^2\theta\,d\phi^2),
\end{align}
with
\begin{align}
S(r,t)=&\frac{R'(r,t)}{\sqrt{1+2E(r)}}.\label{eq:introS}
\end{align}
The LTB metric reduces to the FLRW metric if one sets $R(r,t)=r\,a(t)$ and $E(r)=-\tfrac12 k\, r^2$.

In Ref.~\cite{VanAcoleyen:2008cy} it was shown that when peculiar velocities are small, a seemingly non-linear solution to the metric becomes a linear perturbation on the FLRW metric in the Newtonian gauge.
 Here we focus on solutions to the metric in full generality.

Spherical collapse is studied for example for the formation of black holes as well as for determining, in cosmology, if and when an initially linear over-density produced during inflation collapses and decouples from the background expansion. In simplest approximation, one considers a homogeneous overclosed patch that expands and collapses, matched to an expanding background by a singular shell~\cite{Gunn:1972sv}. Choosing a continuous curvature profile in Eq.~\eqref{eq:introS}, 
allows for an exact solution without singular shells. Spherical collapse in either the approximation or the exact approach, gives insight in clustering of matter, and thereby has been related to presence of for example Dark Energy, 
amongst other possibilities~\cite{Lahav:1991wc,Jhingan:1996jb,Wang:1998gt,Sheth:1999su,Deshingkar:2000hd,Krasinski:2001yi,Krasinski:2003cx,Krasinski:2003yp,Mota:2004pa,Bartelmann:2005fc,Bolejko:2006vw,Firouzjaee:2008gs,Bjaelde:2010qp,Firouzjaee:2011dn}. One should note that initial velocities could be such that an overdensity evolves to become under-dense, but such a decaying mode corresponds to an inhomogeneous Big Bang, which is at tension with the inflationary paradigm in today's favoured model of cosmology.

The formation of voids is studied for two reasons. One reason is the role of voids in the process of structure formation~\cite{Sheth:2003py,Bolejko:2004vb,Martino:2009sf,Biswas:2010ey}, the other is the effect that unusually deep under-densities can have on our perception of Dark Energy. Some studies consider one large local void (size varying from tens to thousands of megaparsecs)~\cite{PhysRevD.45.3512:1992,1995ApJ...453...17M:1995,Tomita:1999qn,Tomita:2001gh,Moffat:2005ii,Mansouri:2005rf,Alnes:2005rw,Moffat:2005yx,Enqvist:2006cg,Chung:2006xh,Enqvist:2007vb,GarciaBellido:2008nz,Zibin:2008vk,Romano:2009ej,Clifton:2009kx,February:2009pv,Sollerman:2009yu,Yoo:2010ad,Clarkson:2010ej,Marra:2010pg,Amendola:2010ub,Biswas:2010xm,Nadathur:2010zm,Alonso:2010zv,Sinclair:2010sb,Marra:2011ct}, where others consider a distribution of many voids, the so called Swiss-Cheese universe~\cite{Marra:2007pm,Biswas:2007gi,Valkenburg:2009iw}.

In presence of only dust and curvature, analytical solutions to the LTB-equation (which we write down later) for $t(R,r)$, {\em i.e.} time as a function of local angular scale factor $R$ and coordinate radius $r$, are known in terms of hyperbolic functions (which become trigonometric functions in case of complex arguments). However, observations of distant Supernovae, of the Baryonic Acoustic Oscillations and of the distance to the surface of last scattering of Cosmic Microwave Background photons combined with locally observed expansion rate, demand the presence of a cosmological constant~\cite{Kessler:2009ys,Percival:2009xn,Komatsu:2010fb,Sandage:2006cv}. It is crucial to realize, although it is not the topic of this paper, that the presence of the cosmological constant is only then necessary to explain the observed geometrical distances, if one assumes that on large enough scales the universe is still properly described by the FLRW metric {\em and} one assumes that the angular diameter distance-redshift relation is correctly described by the background dynamics only.

In the presence of a cosmological constant, dust and curavture, the solution for $t(R,r)$ is an elliptic integral. Therefore, in most works, authors elude numerical integration of the Einstein equations, although see Refs.~\cite{Romano:2010nc,Romano:2011mx}. However, if one for example wants to solve geodesic equations, one typically would perform a numerical integration of the geodesic equations over a numerical solution of the background. The unknown error in the numerical background solution can propagate into the solution to the geodesic equations, possibly leading to unreliable answers, or very slow and sometimes unstable codes. 

In absence of a cosmological constant, but in presence of dust and curvature, the solution for $t(R,r)$ with positive $\kappa(r)$ is,
\begin{align}
 t(R,r)  \propto &  \kappa(r)^{-\frac{3}{2}}  \left\{ \sqrt{\frac{ R \,\kappa(r)}{6r}}\sqrt{3 \frac{R}{r}\, \kappa(r) + 4\pi}-\frac{2^{\frac{3}{2}}\pi}{3}\sinh ^{-1}\left(\sqrt{\frac{3 R \, \kappa(r)}{4\pi r}}\right)\right\},\nonumber
\end{align}
while for $\kappa(r)<0$, the solution becomes a trigonometric in stead of hyperbolic function if one propagates the sign of $\kappa(r)$ correctly.
In this case, one obtains the inverse solution $R(r,t)$ by numerically inverting $t(R,r)$, which can be done quickly and at ultimate accuracy, since $\frac{dt}{dR}$ is known by the definition of the Einstein equations. The thus obtained solution is accurate, fast, and allows for reliable and fast integration of geodesic equations~\cite{Biswas:2010xm}.

The purpose of this paper is to provide all solutions to the background equations $t(R,r)$ that have an initial singularity (Big Bang), in presence of dust, curvature and a cosmological constant, in terms of Elliptic Integrals in Carlson's symmetric form, which can be numerically evaluated as accurate and fast as any elementary function. Compared to the known exact solution in the case of no cosmological constant, the solutions presented in this paper are exact, since one eventually obtains $R(r,t)$ by quick and reliable numerical inversion of $t(R,r)$. Hence, throughout this paper no numerical integration is performed, and solutions are exact but only semi-analytical. See for comparison elliptic solutions involving the Weierstrass elliptic function in Refs.~\cite{1965PNAS...53....1O,1969JETP...29.1027R,Barrow:1984zz,AndrzejKrasinski:1997zz,2003esef.book.....S} and references therein. Note that these references only give $t(R,r)$, which in the FLRW case is enough to solve for $a(t)$, but which in the LTB case does not suffice: $t(R,t)$ only straightforwardly leads to the angular scale factor $R(r,t)$, but we also present for the first time the radial scale factor $S(r,t)$, which is more involved as we shall see later.

The main improvement in this work in comparison with the existing literature, is the fact that we  provide solutions for all functions appearing in the LTB metric, involving spatial derivatives $\partial_r R(r,t)$, necessary for solving for example geodesic equations. We list the solutions for the limits where the local expansion transits to collapse, while a neighbouring shell continues to experience expansion, at the same time preserving continuity of all functions. Moreover, in the appendix we present linear expansions when either the cosmological constant or curvature is small, or both are small. 

The solutions presented in this work allow for a plethora of applications. Let us list but a few. For example, the solutions can be used for:
\begin{itemize}
\item any numerical work involving a general LCDM background expansion,
\item obtaining the exact metric around a collapsing structure
\item simulating the universe as observed by an observer in a large and deep void, in presence of a cosmological constant, allowing for a direct face off between the cosmological constant and the void,
\item studying the evolution of voids in structure formation, in a $\Lambda$CDM universe.
\end{itemize}
We  release a numerical module, written in {\sc Fortran}, that computes exact metric functions and derivatives for a given curvature profile, that can be easily implemented in any code, at  {\myurl}. A brief example of how to invoke this module is given in Appendix~\ref{app:code}.

This work is organized as follows. In Section~\ref{sec:carlson} we first list for reference the used Elliptic Integrals in Carlson's symmetric form. In Section~\ref{sec:metrics} the LTB metric and its Einstein equations are discussed. Then in Section~\ref{sec:soln1} we present one of the main results of this paper, being $t(a)$ in terms of Carlson's elliptic integrals. Next, in Section~\ref{sec:metricfunctions} we solve analytically for the functions in the metric as a function of time $t$ and scale factor $a$. In Section~\ref{sec:examp} we provide an example of an application of the solutions presented in this work. Finally we conclude in Section~\ref{sec:conc}. In Appendix~\ref{app:expansion} we provide the asymptotic expansions of the solutions and in Appendix~\ref{app:radsoln} we show the solution of the Friedmann equation in presence of radiation, matter, curvature and a cosmological constant in terms of Carlson's elliptic integrals.

Throughout this work, square roots of real quantities are taken to be positive, and for all fractional powers of complex numbers $x$ we take the principle value of $\exp(\ln x)$. Extra minus signs due to the possible crossing of branch cuts are written explicitly. We use units in which $G_N=c=1$. Overdots denote time derivatives, primes denote radial derivatives. Our notation follows mostly the notation used in for example Refs.~\cite{Biswas:2007gi,Alexander:2007xx,Biswas:2010xm}.

\section{Carlson's symmetric form of Elliptic Integrals\label{sec:carlson}}
Before discussing solutions to the LTB equation, let us list some definitions for completeness. We take the following definitions of Carlson's symmetric form of elliptic integrals from Ref.~\cite{NIST},
\begin{align}
s(t)=&\sqrt{t+x}\sqrt{t+y}\sqrt{t+z}.\\
\mathop{R_{F}\/}\nolimits\!\left(x,y,z\right)=&\frac{1}{2}\int _{0}^{{\infty}}\frac{dt}{s(t)},\\
\mathop{R_{C}\/}\nolimits\!\left(x,y\right)=&\mathop{R_{F}\/}\nolimits\!\left(x,y,y\right),\\
\mathop{R_{J}\/}\nolimits\!\left(x,y,z,p\right)=&\frac{3}{2}\int _{0}^{{\infty}}\frac{dt}{s(t)(t+p)},\\
\mathop{R_{D}\/}\nolimits\!\left(x,y,z\right)=&\mathop{R_{J}\/}\nolimits\!\left(x,y,z,z\right)=\frac{3}{2}\int _{0}^{{\infty}}\frac{dt}{s(t)(t+z)},\\
3(xyz)^{-\frac{1}{2}}=&R_D(x,y,z)+R_D(z,x,y)+R_D(y,z,x),\label{eq:carlson_relation}
\end{align}
which are defined for $\left\{x,y,z\right\}\in \mathbb{C}\, | \left\{x,y,z\right\}\notin \left<-\infty,0\right]$. These can be evaluated using an iterative procedure, up to unlimited accuracy in very few steps, as explained in Ref.~\cite{1995NuAlg..10...13C}. The definitions are valid for complex arguments, and in all these cases at most one argument is allowed to be zero.

\section{Robertson-Walker and Lema\^itre-Tolman-Bondi metrics\label{sec:metrics}}
The Einstein equations for the FLRW metric and the LTB metric can be written in the same form, with the difference that in the former case no other coordinate dependence than time dependence is present, and in the latter case the curvature and scale factor are both radius and time dependent and radiation is absent. 
The FLRW metric is,
\begin{align}
ds^2 = -dt^2 + a(t) d\vec x^2,
\end{align}
and the Friedmann equation for the FLRW metric is
\begin{align}
 \left( \frac{\dot a}{a} \right)^2 = & H^2(t)=H_0^2 \left[   \frac{\Omega_r}{a^4(t)} + \frac{\Omega_m}{a^3(t)} +  \frac{\Omega_k}{a^2(t)}  + {\Omega_\Lambda}    \right],
\end{align}
where as usual we define today by $t=t_0$, $a_0=a(t_0)=1$, $H_0=H(t_0)$. The different components and their relative abundances are radiation $\Omega_r$,  dust $\Omega_m$, curvature $\Omega_k$ and the cosmological constant $\Omega_\Lambda$.

The LTB metric is given by
\begin{align}
ds^2=-dt^2+S^2(r,t)dr^2+R^2(r,t)(d\theta^2+\sin^2\theta\,d\phi^2),
\end{align}
where
\begin{align}
S(r,t)=&\frac{R'(r,t)}{\sqrt{1+2r^2\kappa(r)\Mtsq}},\\
R(r,t)=&r\,a(r,t).
\end{align}
where ${\Mtsq}$ is an arbitrary parameter defining the length and mass scales, combined with the choice of units $G_{\rm N}=c=1$. 
The Einstein equation leads to the LTB equation~\cite{Bondi:1947av},
\begin{align}
\left( \frac{\dot R}{R} \right)^2 =  \left( \frac{\dot a}{a} \right)^2 = & H^2(r,t)=  \frac{2L(r)}{R^3(r,t)} +  \frac{2r^2\kappa(r)\Mtsq}{ R^2(r,t)}  + \frac{\Lambda}{3} ,\nonumber
\end{align}
where $L(r)= \int_0^r dr\,M'(r) \sqrt{1+2r^2\kappa(r)\Mtsq}$ with $M(r)=4\pi\int_0^r dr\,S(r,t) R^2(r,t) \rho(r,t)$ being the total mass\footnote{
$L(r)$ is the {\em active gravitational mass}, while $M(r)$ is the total rest mass. However, in cosmologically relevant scenarios $2r^2\kappa(r)\Mtsq \ll 1$, such that $L(r)\simeq M(r)$.}  inside radius $r$ and $\rho(r,t)$ is the local matter density.
Now there are three functions of $r$ that specify the problem, $M(r)$, $\kappa(r)$ and $t_{BB}(r)$, where the latter is the radially dependent Big Bang time, $t_{BB}(r)\equiv t(a=0,r)$. One of these three can be fixed to an arbitrary function by redefining the coordinate $r\rightarrow \tilde r=f(r)$ for some monotonic function $f(r)$, without changing the physical description. As discussed in Ref.~\cite{Bolejko:2011ys}, none of the three possible coordinate gauges where one of the functions is fixed to an arbitrary monotonic function captures all possible configurations. In this work we choose the gauge as follows. Demanding a strictly positive matter density, we have $L'(r)\ge 0$. We choose $L(r)=4\pi\Mtsq r^3/3$, such that $L'(r)> 0$ but $L'(r)\neq 0$. This implies that are are no vacuum regions. Then we have 
\begin{align}
M'(r)=&\frac{4\pi\Mtsq r^2}{\sqrt{1+2r^2\kappa(r)\Mtsq}},\\
\rho(r,t)=&\frac{\Mtsq r^2}{R'(r,t)R^2(r,t)}.
\end{align}
In this coordinate gauge the LTB equation becomes,
\begin{align}
\left( \frac{\dot R}{R} \right)^2 =  \left( \frac{\dot a}{a} \right)^2 = & H^2(r,t)=\frac{8 \pi \Mtsq}{3}\left[   \frac{1}{a^3(r,t)} +  \frac{3\kappa(r)}{4\pi a^2(r,t)}  + \frac{\Lambda}{8\pi\Mtsq}    \right],\label{eq:LTB}
\end{align}
and the configuration is completely specified by the two functions $\kappa(r)$ and $t_{BB}(r)$.
The shortcoming of this gauge is, as mentioned above, that it does not allow for solutions with true vacuum over a non-zero range in $r$, for which $M'(r)=0$ such that $\kappa(r)\rightarrow \infty$.

\subsection{Normalization}
Normalizing $a(r_*,t_0)=1$ at a chosen $\{r_*,t_0\}$, we have
\begin{align}
\kappa_b\equiv&\frac{4\pi}{3}\frac{\Omega_\kappa(r_*)}{1-\Omega_\kappa(r_*)-\Omega_\Lambda(r_*)}\label{eq:norm1}\\
\Mtsq \equiv& \frac{3 H^2(r_*,t_0) - {\Lambda}}{8\pi} \frac{1}{1+\frac{3\kappa_b}{4\pi}}, \\
t_0\equiv& t(a(r_*,t_0)) ,
\end{align}
which is regular for $\Lambda \rightarrow 0$ and $\Omega_k \rightarrow 0$. Also, we  write $H_*\equiv H^2(r_*,t_0)$. One can choose an arbitrary $r_*$ at which to normalize, but one has to fix it once and for all. Clearly, in the FLRW case, the choice of $r_*$ is irrelevant since the $r$-depence of $H(r,t)$ vanishes, and we find $H_0=H_*$.

\subsection{Towards solving for $t(a,r)$}
One can define
\begin{align}
\Omega_{m}(r) &=  \frac{8\pi\Mtsq}{3 H^2(r,t_0)},\label{eq:om_mat_r}\\
\Omega_{k}(r) &=  \frac{2 \Mtsq}{ H^2(r,t_0)} \kappa(r),\label{eq:om_k_r}\\
\Omega_{\Lambda}(r) &=  \frac{1}{ H^2(r,t_0)}\frac{ \Lambda  }{  3},\label{eq:om_lambda_r}
\end{align}
in which case one retrieves the Friedmann equation in absence of radiation when one drops the $r$-dependences in equation~\eqref{eq:LTB} and when one normalizes $a(r,t_0)=a(t_0)=1$. For a generic matter distribution in the LTB metric one however has $a(r,t_0)\neq 1$. Then, in the LTB metric, these three quantities are the relative content at $a(t,r)=1$ in a shell at a given radius: dust, curvature and cosmological constant respectively. At $t_0$ the relative matter quantities are then depending on the value of $a(r,t_0)$, {\em e.g.} $\Omega_m a^{-3}(r,t_0)$ for matter. The reader should note that in this sign convention 
\begin{itemize}
\item {\color{mydarkdarkblue} {$\mathbf\Omega_k>0$}} corresponds to an {\color{mydarkdarkblue}{\bf open universe}} 
\item and vice versa {\color{mydarkdarkred} {$\mathbf\Omega_k<0$}} corresponds to a {\color{mydarkdarkred} {\bf closed universe}}. 
\end{itemize}

From this point on we will neglect radiation, although the reader is referred to Appendix~\ref{app:radsoln} for a discussion of the solution in presence of radiation, only valid in the FLRW metric.

Writing $A\equiv a(r,t)$, the general solution to $t$ for the metric (both LTB and FLRW) is given by the integral,
\begin{align}
\int_0^{A}&\frac{\sqrt{\tilde a}\,d\tilde a}{\sqrt{ {\Omega_m}(r) +  {\Omega_k}(r){\tilde a}  + {\Omega_\Lambda}(r) {\tilde a^3 }}}=H_*\left[t(A) - t_{BB}(r)\right],\label{eq:gen_tofa}
\end{align}
where we take the positive square root and the integral is performed at constant $r$. Note that the Big Bang time $t_{BB}(r)$ acts as an integration constant for the left hand side of this equation.

In the case of existence of at least one positive real root of the polynomial in the denominator ({\em i.e.} an over-closed universe (FLRW) or over-closed shell at radius $r$ (LTB)), there are two solutions for $t(A)$, one for the expanding and one for the collapsing phase. For example labeling the smallest positive real root $U(r)$, such that the turning point in the expansion history lies at $a(r,t)=U(r)$, the second (collapsing) solution is
\begin{align}
H_*&\left[t(A) - t_{BB}(r)\right]=\nonumber\\
&\int_0^{U(r)}\frac{\sqrt{\tilde a}\,d\tilde a}{\sqrt{ {\Omega_m}(r) +  {\Omega_k}(r){\tilde a}  + {\Omega_\Lambda}(r) {\tilde a^3 }}}\mp\int_A^{U(r)}\frac{\sqrt{\tilde a}\,d\tilde a}{\sqrt{ {\Omega_m}(r) +  {\Omega_k}(r){\tilde a}  + {\Omega_\Lambda}(r) {\tilde a^3 }}},\label{eq:gen_tofa_ex}
\end{align}
where the sign in front of the second integral is determined by whether $\dot a$ changes sign (collapse) or not (continued expansion) at $a=U$. Throughout the rest of this work, we discard cases where $\dot a$ does not change sign, {\em i.e.}, we only consider cases where $\ddot a$ is non-zero at $a=U(r)$. The case with more than one positive real root then becomes irrelevant: when $\dot a$ changes sign, by symmetry the contraction is identical to the expansion, and we only consider the branch of solutions that experiences an initial singularity (Big Bang). As $a$ shrinks again towards $0$, for each value of $a$ the time derivative is exactly $-1$ times the time derivative for the same value of $a$ during the expansion. The higher roots are never encountered.

\section{Solving for the expansion rate\label{sec:soln1}}
We rewrite  Eq.~\eqref{eq:gen_tofa} to 
\begin{align}
H_* \left[t(A) - t_{BB}(r)\right]=&\frac{1}{\sqrt{\Omega_\Lambda}}\int_0^{A} \frac{a^{\frac{n-2}{2}} d\,a}{\sqrt{\prod_{m=1}^n (a-y_m)}} ,\label{eq:tofa_prod_0_A}
\end{align}
for $n=3$, where we rewrote the polynomial
\begin{align}
 \Omega_m(r) +  \Omega_\kappa(r) a  + \Omega_\Lambda(r)  a^3 = \Omega_\Lambda (a-y_1)(a-y_2)(a-y_3),
 \end{align}
with $y_i$ the solutions to  
\begin{align}
  \frac{\Omega_m(r)}{\Omega_\Lambda(r)} +  \frac{\Omega_\kappa(r)}{\Omega_\Lambda(r)} y_i  +  y_i^3 = 0.
 \end{align}

We now turn Eq.~\eqref{eq:tofa_prod_0_A} into Carlson's symmetric form by subsequently making the change of variables $a\rightarrow c=\frac{1}{a}$ and $c\rightarrow b=c-\frac{1}{A}$, such that we get
\begin{align}
H_* \left[t(A) - t_{BB}(r)\right]=&\frac{1}{\sqrt{\Omega_\Lambda}}\int_0^{A} \frac{a^{\frac{n-2}{2}} d\,a}{\sqrt{\prod_{m=1}^n (a-y_m)}}\nonumber\\
=&\frac{1}{\sqrt{\Omega_\Lambda}}\int_{\frac{1}{A}}^{\infty} \frac{c^{-\frac{n-2}{2}-2} d\,c}{\sqrt{\prod_{m=1}^n (\frac{1}{c}-y_m)}}\nonumber\\
=&\frac{1}{\sqrt{\Omega_\Lambda}}\frac{(-1)^{-\frac{3n}{2}}}{\sqrt{\prod_{m=1}^n y_m}} \int_{\frac{1}{A}}^{\infty} \frac{d\,c}{c\sqrt{\prod_{m=1}^n (c-\frac{1}{y_m})}}\nonumber\\
=&\frac{1}{\sqrt{\Omega_\Lambda}}\frac{(-1)^{-\frac{3n}{2}}}{\sqrt{\prod_{m=1}^n y_m}} \int_{0}^{\infty} \frac{d\,b}{(b+\frac{1}{A})\sqrt{\prod_{m=1}^n (b+\frac{1}{A}-\frac{1}{y_m})}}.\label{eq:gensoln}
\end{align}
These transformations are valid, since no rotations in the complex plane are involved, and none of the roots is transformed onto the path of integration; no branch points come to lie on the positive real axis. At most one root will be at the zero of the integration domain in the variable $b$, when we integrate $\int_0^{y_m}d\,a$. This exception is allowed in the definition of the symmetric elliptic integrals in Section~\ref{sec:carlson}.

In terms of physics this is straight forward to see: for any real negative $y_m$, we have $\frac{1}{A}-\frac{1}{y_m}>0$, {\em i.e.} the branch point lies on the negative real axis of $b$; for any real positive $y_m$, we also have $\frac{1}{A}-\frac{1}{y_m}\geq0$, since we integrate at most up to $A=y_m$. The scale factor never grows beyond its smallest maximum for $t\in \mathbb{R}$.

We now obtained the solution for the time as a function of scale factor expressed as a symmetric elliptic integral in Eq.~\eqref{eq:gensoln}, that is,
\begin{align}
H_* \left[t(A) - t_{BB}(r)\right]=&\frac{2}{3\sqrt{\Omega_\Lambda}}\frac{(-1)^{-\frac{9}{2}}}{\sqrt{\prod_{m=1}^3 y_m}} R_J\left(\frac{1}{A}-\frac{1}{y_1},\frac{1}{A}-\frac{1}{y_2},\frac{1}{A}-\frac{1}{y_3}, \frac{1}{A}  \right).
\end{align}
One of the first three arguments of $R_J(x,y,z,p)$ is allowed to be zero, such that the limit of $A\rightarrow y_1$ for positive real $y_1$ is trivial.

In order to keep a connection to the well-known Friedmann equation, we so far used a notation in terms of $H_*$ and $\Omega_i(r)$. However, as $H_*=H(r,t_0)$ is a function of $r$, it is more convenient to go back to the original form of Eq.~\eqref{eq:LTB} and recast Eq.~\eqref{eq:gen_tofa} as
\begin{align}
t(A) - t_{BB}(r) =&\frac{1}{\sqrtMtsq}\int_0^{A}\frac{\sqrt{\tilde a}\,d\tilde a}{\sqrt{ \frac{8 \pi}{3} +  2 \kappa(r) {\tilde a}  + \frac{\Lambda}{3\Mtsq}  {\tilde a^3 }}},\label{eq:t_LTB_full}\\
=&\frac{2}{\sqrt{3\Lambda}}\frac{(-1)^{-\frac{9}{2}}}{\sqrt{\prod_{m=1}^3 z_m}} R_J\left(\frac{1}{A}-\frac{1}{z_1},\frac{1}{A}-\frac{1}{z_2},\frac{1}{A}-\frac{1}{z_3}, \frac{1}{A}  \right),\label{eq:t_LTB_full_soln}
\end{align}
where the roots $y_i$ and $z_i$ are related by a simple rescaling.

\subsection{The roots $z_i$}
The roots $z_i$ are trivial,
\begin{align}
z_1=&\frac{{\Phi}^{\frac{1}{3}}}{{2}^{\frac{1}{3}} 3^{\frac23} Z}-\frac{{\left(\frac{2}{3}\right)^{\frac{1}{3}}} Y}{{\Phi}^{\frac{1}{3}}},\\
z_2=&\frac{\left(1+i \sqrt{3}\right) Y}{2^{\frac23} {3}^{\frac{1}{3}} {\Phi}^{\frac{1}{3}}}-\frac{\left(1-i \sqrt{3}\right)
   {\Phi}^{\frac{1}{3}}}{2^{\frac{4}{3}} 3^{\frac23} Z},\\
z_3=&\frac{\left(1-i \sqrt{3}\right) Y}{2^{\frac23} {3}^{\frac{1}{3}} {\Phi}^{\frac{1}{3}}}-\frac{\left(1+i \sqrt{3}\right)
   {\Phi}^{\frac{1}{3}}}{2^{\frac{4}{3}} 3^{\frac23} Z},
\end{align}
where $\Phi=\sqrt{3} \sqrt{27 X^2 Z^4+4 Y^3 Z^3}-9 X Z^2$, and $X=\frac{8\pi}{3}$, $Y=2\kappa(r)$ and $Z=\frac{\Lambda}{3\Mtsq}$. For the FLRW metric the roots $y_i$ obey the same expressions, replacing $X$, $Y$ and $Z$ by the corresponding $\Omega_i$.

\section{The metric functions and their time derivatives\label{sec:metricfunctions}}
In this section we present the main result of this paper, the radial scale factor and radial derivatives of other metric functions. In the previous section we discussed the solution $t(a)$. Since we know by definition the exact $\frac{dt}{da}$, one can solve numerically for $a(t)$ using a simple Newton-Raphson algorithm, obtaining $a(t)$ at machine accuracy level at hardly any computational cost. Therefore, in the following we assume $\{r,t\}$ as input parameters, and the function $a(r,t)$ as a known function. We aim to express all solutions in terms of those quantities.

The functions appearing in the metric are $R(r,t)$ and $R'(r,t)$. The time derivatives of these functions are relevant when one wants to solve geodesic equations in this metric, which is why we discuss them here as well.  We have $R(r,t)\equiv r a(r,t)$, such that 
\newlength{\mystretch}
\setlength{\mystretch}{.3em}
\begin{align}
R'(r,t)=&a(r,t) + r a'(r,t)\\[\mystretch]
\dot R(r,t) =& r \dot a(r,t)\\[\mystretch]
\dot R'(r,t)=& \dot a(r,t) + r \dot a'(r,t)\nonumber\\[\mystretch]
=&\dot a + r a'(r,t) H(r,t) + r a(r,t) H'(r,t)\label{eq:rpd1}\\[\mystretch]
H'(r,t)=&\frac{1}{2H(r,t)}\partial_rH^2(r,t)\nonumber\\[\mystretch]
=&\frac{1}{2H(r,t)}\left(     \frac{8 \pi \Mtsq}{3}\left[   \frac{-3}{a^4(r,t)} +  \frac{-3\kappa(r)}{2\pi a^3(r,t)}     \right] a'(r,t)  +  \frac{3\kappa'(r)}{4\pi a^2(r,t)} \right).\label{eq:hprime1}
\end{align}
Comparing Eq.~\eqref{eq:rpd1} to Eq.~\eqref{eq:hprime1}, we see that we only need the solution for $a'(r,t)$ in order to be able to calculate all relevant quantities. However, the term $\frac{1}{2H(r,t)}$ asks for care to be taken at the transition from expansion to collapse, where $H(r,t)=0$. We will show in the following that this limit is in fact regular, and that all metric functions remain properly defined throughout all the expansion and collapse history.

\subsection{Spatial derivative of the scale factor during expansion\label{sec:spat_der_full}}
Since $t$ is one of the orthogonal coordinates, we have $\partial_r t \equiv0$, even if we solve for $t$ by $t=t(a(r,t),r)$. Therefore,
\begin{align}
\frac{d}{dr}\left[t(a,r)-t_{BB}(r)\right]=&a'(r,t)\frac{\partial}{\partial a} t(a,r)+\frac{\partial}{\partial r}\left[t(a,r)-t_{BB}(r)\right]\nonumber\\
=0-\partial_r t_{BB}(r)= &a'(r,t)\frac{1}{\dot a(r,t)}+\frac{\partial}{\partial r}\left[t(a,r)-t_{BB}(r)\right]\\
a'(r,t)=&-\dot a(r,t) \left[  \frac{\partial}{\partial r}\left[t(a,r)-t_{BB}(r)\right]+\partial_r t_{BB}(r)  \right].\label{eq:spatial_firstpart}
\end{align}
The only non-trivial term in~\eqref{eq:spatial_firstpart} is 
\begin{align}
\sqrtMtsq \frac{\partial}{\partial r}\left[t(a,r)-t_{BB}(r)\right] =&  \frac{\partial}{\partial r}\int_0^{A}\frac{\sqrt{a}\,d a}{\sqrt{{X}+Y\,a+Z\, a^3}}\nonumber\\
 =&-\frac{Y'}{2}\int_0^{A}\left[\frac{a}{{{X}+Y\,a+Z\, a^3}}\right]^{\frac{3}{2}}d\,a,
\end{align}
where we  continue to use the notation of the previous section, $A=a(r,t)$, $X=\frac{8\pi}{3}$, $Y=2\kappa(r)$ and $Z=\frac{\Lambda}{3\Mtsq}$. This expression  is again an elliptic integral. We spare the reader the detailed steps, but the general procedure is very much like in Eq~\eqref{eq:gensoln}, however one not only substitutes 
$a\rightarrow c=\frac{1}{a}$ and $c\rightarrow b=c-\frac{1}{A}$, but one also splits into partial fractions, to arrive at 
\begin{align}
\sqrtMtsq \frac{\partial}{\partial r}\left[t(A,r)-t_{BB}(r)\right] =&-\frac{Y'}{2Z^{\frac{3}{2}}}\int_0^{A}\left[\frac{a}{{\frac{X}{Z}+\frac{Y}{Z}\,a+ a^3}}\right]^{\frac{3}{2}}d\,a\nonumber\\
=&\frac{Y'}{3Z^{\frac{3}{2}}}\frac{i}{( z_1z_2z_3)^{\frac{1}{2}}}\left[\frac{1}{(z_1-z_2)(z_1-z_3)}R_D\left(\frac{1}{A}-\frac{1}{z_2}, \frac{1}{A}-\frac{1}{z_3},\frac{1}{A}-\frac{1}{z_1}\right) \right. \nonumber\\
&+\frac{1}{(z_2-z_1)(z_2-z_3)}R_D\left(\frac{1}{A}-\frac{1}{z_3}, \frac{1}{A}-\frac{1}{z_1},\frac{1}{A}-\frac{1}{z_2}\right)  \nonumber\\
&\left.+\frac{1}{(z_3-z_1)(z_3-z_2)}R_D\left(\frac{1}{A}-\frac{1}{z_1}, \frac{1}{A}-\frac{1}{z_2},\frac{1}{A}-\frac{1}{z_3}\right) \right] \label{eq:symm_spat_deriv} .
\end{align}
One can take this equation one step further, using Eq.~\eqref{eq:carlson_relation}, which is $R_D(x,y,z)+R_D(z,x,y)+R_D(y,z,x)=3(xyz)^{-\frac{1}{2}}$. Hereby one eliminates one evaluation of the function $R_D(x,y,z)$, and more importantly, it reveals the kind of singularity that is encountered when one of the arguments goes to zero. Choosing $z_1$ to be the smallest positive real root, or as in the previous section $U(r)=z_1$, we finally have the solution,
\begin{align}
& \sqrtMtsq  a'(r,t)=  -\frac{Y'}{Z} \frac{A}{(z_1-z_2)(z_1-z_3)} - \frac{\sqrt{Z}\sqrt{(A-z_1)(A-z_2)(A-z_3)}}{\sqrt{A}}  \times \nonumber\\
&  \left[ \frac{Y'}{3Z^{\frac{3}{2}}}\frac{i}{( z_1z_2z_3)^{\frac{1}{2}}} \left(\frac{1}{(z_2-z_1)(z_2-z_3)}-\frac{1}{(z_1-z_2)(z_1-z_3)}\right)R_D\left(\frac{1}{A}-\frac{1}{z_3}, \frac{1}{A}-\frac{1}{z_1},\frac{1}{A}-\frac{1}{z_2}\right) \right.  \nonumber\\
&\left. +  \frac{Y'}{3Z^{\frac{3}{2}}}\frac{i}{( z_1z_2z_3)^{\frac{1}{2}}} \left(\frac{1}{(z_3-z_1)(z_3-z_2)}-\frac{1}{(z_1-z_2)(z_1-z_3)}\right)R_D\left( \frac{1}{A}-\frac{1}{z_1}, \frac{1}{A}-\frac{1}{z_2},\frac{1}{A}-\frac{1}{z_3}\right)\right.\nonumber\\
&\left.+\sqrtMtsq\partial_r t_{BB}(r)  \phantom{\int}\right],
\end{align}
with $A=a(r,t)$, which can be recast as
\begin{align}
&\sqrtMtsq a'(r,t) =\dot a(r,t) Q(r,t) + P(r,t),\label{eq:aprime_final}\\
&P(r,t) =    -\frac{Y'}{Z} \frac{A}{(z_1-z_2)(z_1-z_3)} \label{eq:aprime_final_P}\\
&Q(r,t)=-\left[\sqrtMtsq\partial_r t_{BB}(r)  \phantom{\int}\right.\nonumber\\
& \left. +  \frac{Y'}{3Z^{\frac{3}{2}}}\frac{i}{( z_1z_2z_3)^{\frac{1}{2}}} \left(\frac{1}{(z_2-z_1)(z_2-z_3)}-\frac{1}{(z_1-z_2)(z_1-z_3)}\right)R_D\left(\frac{1}{A}-\frac{1}{z_3}, \frac{1}{A}-\frac{1}{z_1},\frac{1}{A}-\frac{1}{z_2}\right) \right.  \nonumber\\
&\left. + \frac{Y'}{3Z^{\frac{3}{2}}}\frac{i}{( z_1z_2z_3)^{\frac{1}{2}}} \left(\frac{1}{(z_3-z_1)(z_3-z_2)}-\frac{1}{(z_1-z_2)(z_1-z_3)}\right)R_D\left( \frac{1}{A}-\frac{1}{z_1}, \frac{1}{A}-\frac{1}{z_2},\frac{1}{A}-\frac{1}{z_3}\right)\right],\label{eq:aprime_final_Q}
\end{align}
with as before $A\equiv a(r,t)$.

If we look back at Eq.~\eqref{eq:spatial_firstpart} and Eq.~\eqref{eq:symm_spat_deriv}, we see  that the overall factor $\dot a(r,t)$ in the expression for $a'(r,t)$ multiplies $\dot a^{-1}$  inside $\partial_r \left[ t(a,r) - t_{BB}(r)\right]$, such that $a'(r,t)$ is finite and non-zero for $\dot a(r,t)\rightarrow 0$, that is $\lim_{\dot a(r,t)\rightarrow 0}a'(r,t) = P(r,t)/\sqrtMtsq$.

\subsection{Spatial derivative of the scale factor during collapse}
During the collapsing phase, we have
\begin{align}
\sqrtMtsq \left[ t(A)-t_{BB}(r) \right]=&\frac{2}{\sqrt{Z}}\int_0^{U(r)}\frac{\sqrt{a}\,d a}{\sqrt{(a-z_1)(a-z_2)(a-z_3)}}\nonumber\\ & - \frac{1}{\sqrt{Z}}\int_0^{A}\frac{\sqrt{a}\,d a}{\sqrt{(a-z_1)(a-z_2)(a-z_3)}}.
\end{align}
We write explicitly the $r$-dependence in $U(r)\equiv z_1(r)$, to point out that the transition from expansion to collapse is {\em a priori} an $r$-dependent event.\footnote{Actually, we have $U(r)\equiv a(r,t_U)$, where $t_U$ denotes the time at which $\dot a =0$. Since this time $t_U$ is itself a function of $r$, one can show that $U'(r)=a'(r,t_U) + a(r,t_U)\partial_r t_U = \partial_r z_1$.} Taking the derivative of this expression, we find
\begin{align}
\sqrtMtsq a'(r,t)=& \left| \dot a \right| \left[      \partial_{r} \left\{ \frac{2}{\sqrt{Z}}\int_0^{z_1(r)}\frac{\sqrt{a}\,d a}{\sqrt{(a-z_1)(a-z_2)(a-z_3)}} \right\} \right. \nonumber\\ &
\left. +\frac{Y'}{2}\int_0^{A}\left[\frac{a}{{{X}+Y\,a+Z\, a^3}}\right]^{\frac{3}{2}}d\,a   +\sqrtMtsq \partial_r t_{BB}(r) \right],
\end{align}
where the absolute value $\left|\dot a\right|$ is to remind us that we take the positive root of $\sqrt{(a-z_1)(a-z_2)(a-z_3)}$.
When we realize that the $r$-dependence in $z_1(r)$ is entirely specified by $Y(r)$ as that is the only $r$-dependent function in all integrals, such that $\partial_r z_1=Y'(r) \partial_Y z_1$, the first term becomes,
\begin{align}
 \partial_{r}& \left\{ \frac{2}{\sqrt{Z}}\int_0^{z_1}\frac{\sqrt{a}\,d a}{\sqrt{(a-z_1)(a-z_2)(a-z_3)}}\right\}=\nonumber\\
&\partial_r Y \lim_{A\rightarrow z_1}\left\{ \frac{2\partial_{Y}z_1}{\sqrt{Z}}\frac{\sqrt{A}}{\sqrt{(A-z_1)(A-z_2)(A-z_3)}}  
-  \frac{1}{Z^{\frac{3}{2}}}\int_0^{A}\left(\frac{a}{(a-z_1)(a-z_2)(a-z_3)}\right)^{\frac{3}{2}}\,d a\right\}.
\end{align}
As $z_1$ satisfies $X+Y\,z_1+Z\,z_1^3=0$, we have $\partial_Y z_1 = \frac{-z_1}{Y+3Z\,z_1^2}$. Next, we observe that 
\begin{align}
Y+3Z\,z_1^2=&Z\lim_{B\rightarrow z_1} \partial_{B} \left(\frac{X}{Z} + \frac{Y}{Z} B + B^3\right)\nonumber\\
 = &Z\lim_{B\rightarrow z_1} \partial_{B} (B-z_2)(B-z_3)(B-z_1)\nonumber\\
 = &Z \lim_{B\rightarrow z_1} \left[ (B-z_3)(B-z_1)+(B-z_2)(B-z_1)+(B-z_2)(B-z_3)\right]\nonumber\\
 =&Z(z_1-z_2)(z_1-z_3).\label{eq:zy_uvw}
\end{align}
so that $\partial_Y z_1 = \frac{-z_1}{Z(z_1-z_2)(z_1-z_3)}$. Using this relation, together with the solution to the same integral in Section~\ref{sec:spat_der_full}, and some more simple algebra, we arrive at 
\begin{align}
a_{\rm coll}'(r,t)=\dot a \left(2Q(r,t_U) - Q(r,t)\right) + P(r,t),\label{eq:aprime_collapse_abrev}
\end{align}
where $t_U\equiv t(a=z_1,r)$, the subscript `${\rm coll}$' denotes that this expression is valid during a collapsing phase, $Q(r,t)$ and $P(r,t)$ are defined in Eqs.~(\ref{eq:aprime_final_P},\ref{eq:aprime_final_Q}), and $Q(r,t_U)$ is evaluated by replacing $A\rightarrow z_1$ in Eq.~\eqref{eq:aprime_final_Q}.

It should be understood now that with expressions~\eqref{eq:aprime_final}~and~\eqref{eq:aprime_collapse_abrev}, $a'(r,t)$ is finite and continuous at $\dot a \rightarrow 0$.

\subsection{The spatial derivative of the Hubble parameter at $\dot a =0$}
During expansion, the expression for $H'(r,t)$ in Eq~\eqref{eq:hprime1} is regular. At $\dot a \rightarrow 0$, we can now insert Eq.~\ref{eq:aprime_final} for $a'(r,t)$, to find after some manipulations,
\begin{align}
\sqrtMtsq H'(r,t)=&-\left( \frac{3X}{2a^3}+\frac{Y}{a^2}  \right) Q(r,t) - \frac{3XY'}{2\sqrt{Z} a^{\frac{3}{2}} ({3X+2Yz_1})} \sqrt {\frac{a - z_1}{(a-z_2)(a-z_2)}} \label{eq:Hprime2},
\end{align}
which is pefectly regular for the whole domain $a(r,t) \in \{0,z_1\}$. One extra minus sign appeared in the last expression, following from $\frac{a-z_1}{\sqrt{a-z_1}}=-\sqrt{a-z_1}$ with $a\leq z_1$ for all $a$.

\subsection{The spatial derivative of the Hubble parameter during collapse}
Combining Eq.~\eqref{eq:aprime_collapse_abrev} and Eq.~\eqref{eq:Hprime2}, we find during a collapsing phase,
\begin{align}
H_{\rm coll}'(r,t)=&-\left( \frac{3X}{2a^3}+\frac{Y}{a^2}  \right) \left(2Q(r,t_U) - Q(r,t)\right)  - \frac{3XY'}{2\sqrt{Z} a^{\frac{3}{2}} ({3X+2Yz_1})} \sqrt {\frac{a - z_1}{(a-z_2)(a-z_3)}}, 
\end{align}
preserving continuity at $\dot a = 0$.

\section{Example of application\label{sec:examp}}
\begin{figure}
\begin{center}
\includegraphics[width=0.98\textwidth]{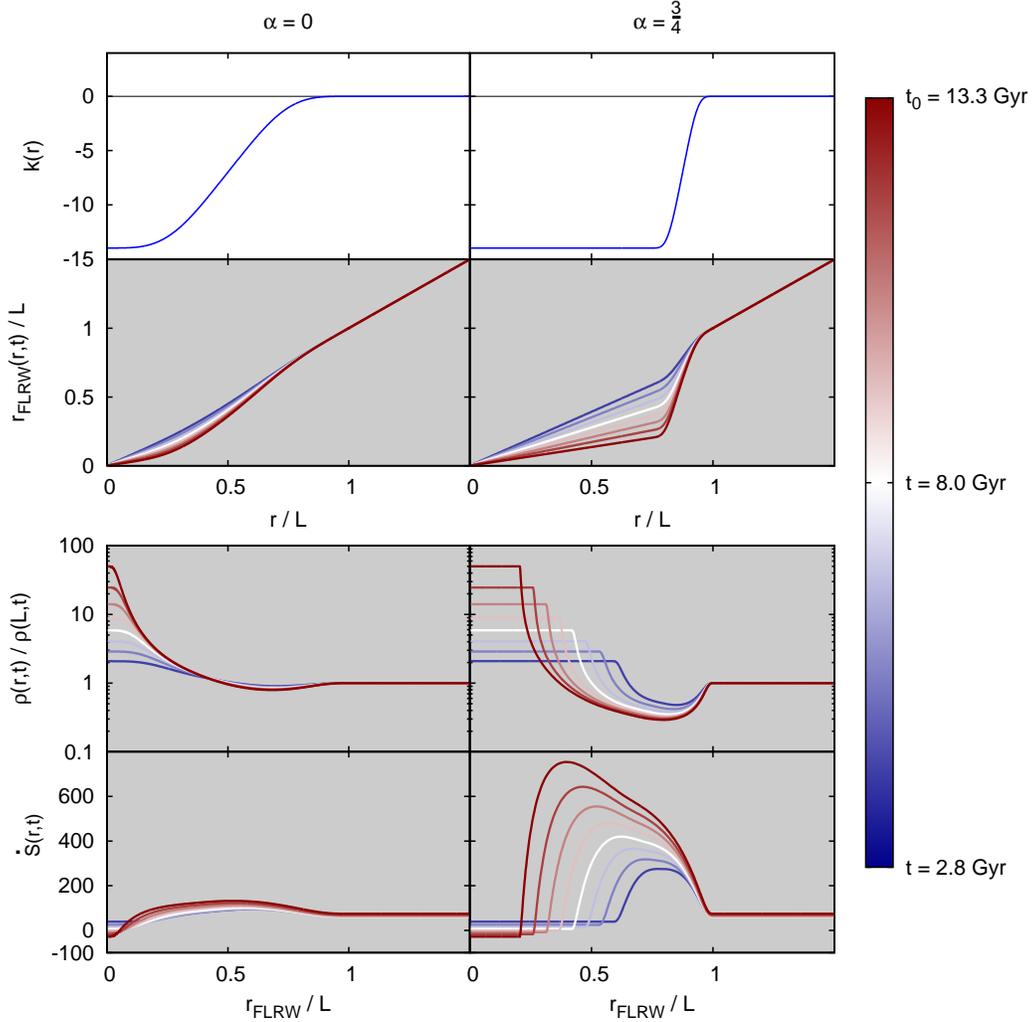}
\end{center}
\caption{Comparison of the two distinct overdensities. Curvature profile (top row) as a function of comoving radius, the auxiliary FLRW radius $r_{\rm FLRW}$ as a function of radius and time (second row), the matter density as a function of the auxiliary radius $r_{\rm FLRW}$ and time (third row) and finally the time derivative of the metric function $S(r,t)$ as a function of auxiliary radius $r_{\rm FLRW}$ and time (bottom row).
 Both over-densities are matched to the same homogeneous $\Lambda$CDM-universe, at the same comoving radius $r=1$ Mpc. The difference between the over-densities is in the value $\alpha$ in Eq.~\eqref{eq:theprofile}, determining the range in $r$ over which the curvature profile falls back to the background value and thereby determining the total mass in the over-density. In all graphs, the time coordinate is represented by the colour of the curves, evolving from 
 red to white to blue (from black to white to black in black-and-white print), 
 indicating time varying from today ($t_0=13.3$ Gyr) to some moment in the past ($t_0=2.8$ Gyr). Additionally, labels inside the graphs indicate times corresponding to different curves.}\label{fig:thefig}
\end{figure}

In Figure~\ref{fig:thefig} we show an example of an application of the solutions presented in this work. The figure shows a comparison between two different density profiles, both parametrized by
\setlength{\mystretch}{1em}
\begin{align}
\kappa(r)=\kappa_{\rm max} W_3\left(\frac{r}{L},\alpha\right)+\kappa_b, 
\end{align}
where $\kappa_b$ is defined in Eq.~\eqref{eq:norm1}, and with
\begin{align}
W_3\left(\frac{r}{L},\alpha\right)=\left\{\begin{array}{lr}
1& \mbox{ for } r < \alpha L\\[\mystretch]
\frac{1}{4\pi^2}\left[ 1 + \pi^2 \left ( 4 - 8 \left(\frac{r - \alpha L}{L-\alpha L}\right)^2 \right) -  \cos \left( 4\pi \frac{r - \alpha L}{L-\alpha L}  \right) \right] & \mbox{ for } \alpha L< r < \frac{1+\alpha}{2} L\\[\mystretch]
\frac{1}{4\pi^2}\left[ - 1 + 8\pi^2 \left ( \frac{r - \alpha L}{L-\alpha L} -1 \right)^2 +  \cos \left( 4\pi \frac{r - \alpha L}{L-\alpha L}  \right) \right] & \mbox{ for }  \frac{1+\alpha}{2} L< r < L\\[\mystretch]
0& \mbox{ for } r\geq L,
\end{array}\right.  \label{eq:theprofile}
\end{align}
which is the third order of the interpolating function $W_n(x,\alpha)$ which interpolates from $1$ to $0$ in the interval $\alpha < x < 1$, while remaining $C^n$ everywhere. We introduce the interpolating function $W_n(x,\alpha)$ in Appendix~\ref{app:wfunc}. 
For $0<\alpha<1$ 
this function is $C^3$ on $r\in \left [ 0,\infty \right >$. Hence, all functions in the metric are $C^2$ everywhere, including at the center ($r=0$) and at the matching to FLRW at $r=L$, guaranteeing a finite and continuous Ricci scalar everywhere. By construction all functions shown in the figure in fact have continuous first derivatives in $r$, even if by eye it may seem otherwise.

The left column in Figure~\ref{fig:thefig} shows quantities for a profile with $\alpha=0$, the right column shows the same quantities for $\alpha=\frac34$. In the top row we see the curvature profiles as a function of comoving radius, which are time independent. Both profiles share a same matter density at the centre where $r=0$ and at the outer radii $r>L$, however the difference in shape of profile leads to a different total amount of matter inside the over-density. For $\alpha=\frac34$, the over-density possesses a larger region with large closed curvature. The second row shows a time-dependent auxiliary radius,
\begin{align}
r_{\rm FLRW}(r,t)\equiv \int_0^{r} \frac{S(r',t) }{ a(\infty,t)} dr',
\end{align}
which keeps the length measure along this radius at a given time constant, $ds^2=S(r,t)dr^2=dr_{\rm FLRW}^2$, but not constant in time. This radius can be interpreted loosely as what an FLRW observer would see. For these scenarios it turns out that roughly $r_{\rm FLRW}\simeq R'(r,t)/a(\infty,t)$, which is its definition in Refs~\cite{Marra:2007pm,Valkenburg:2009iw}.

In the third row we show the relative matter density, normalized to the matter density of the surrounding homogeneous cosmology. While at the centre the time evolution is the same for both profiles, the surrounding under-dense (but still closed curved) shell differs largely between the two cases. Since $\kappa(r)$ stays large and negative up to higher radius in the $\alpha=\frac34$-case, compared to the $\alpha=0$-case, a larger range in $r$ is present for which shells have a collapsing solution and actually experience the collapse. Therefore, the range in $r$ in which the shells expand rapidly and become more and more under-dense, is smaller for the $\alpha=\frac34$-case. Hence, the resulting surrounding under-density is less dense and more emphasized in the $\alpha=\frac34$-case than in the $\alpha=0$-case.

For a radial null geodesic, the trajectory is defined by,
\begin{align}
\frac{dt}{dr} &= - S(r,t)\\
\frac{dz}{dr} &= (1+z)\dot S(r,t).
\end{align}
This is why, for illustration, we plot $\dot S(r,t)$ in the fourth row in Figure~\ref{fig:thefig}.

Obviously, as the effect of the spherical collapse on the outer radii is more violent for the $\alpha=\frac34$-case, the red shift that a photon experiences by passing through that region is larger than in the $\alpha=0$-case. Even though this observation is not enough to draw conclusions about photon geodesics and collapsing structures, it illustrates how the solutions in this work can be used to further asses the importance of the initial distribution of matter in the line of sight, on distant observations.

Note that all quantities remain perfectly smooth and continuous at the transition from expansion to collapse, which in the third and fourth row in Figure~\ref{fig:thefig} occurs roughly where each curve crosses the level of the background, {\em i.e.} the level of the same curve at $r>L$. 

The solutions presented in this work allow for practically instantaneous calculation of the quantities presented in Figure~\ref{fig:thefig}. We release a module, written in {\sc Fortran}, which returns all metric functions and derivatives thereof as a function of time for given functions $\kappa(r)$ and $t_{BB}(r)$, and given cosmological parameters. The module is released at\\ {\myurl}.

\section{Conclusion\label{sec:conc}}
We have presented an as complete as possible overview of the solutions to the Einstein equations governing the Lema\^itre-Tolman-Bondi metric, including fully continuous solutions for collapsing over-densities surrounded by an expanding universe. The solutions are written in terms of Elliptic Integrals in Carlson's symmetric form, which allow for fast numerical evaluation of the solutions at machine accuracy level. The solutions to all metric functions involve the numerical inversion of one function, $t(a)$, whose derivative is explicitly known {\em a priori}, therefore allowing for inversion at machine accuracy level while remaining sufficiently fast.

We finished with a brief example of how these solutions can be applied. These solutions could improve the accuracy and speed of many analyses involving structure formation and inhomogeneous cosmologies.

\section*{Acknowledgements}
The author wishes to thank Yvonne Wong, Tomislav Prokopec and Valerio Marra for useful discussions.

\appendix
\section{Asymptotic expansions\label{app:expansion}}
The solutions of the scenario where all three components, dust, curvature and the cosmological constant are non-zero and not asymptotically small, are given in the main body of this paper. For several purposes, a not unimportant one being numerical accuracy, asymptotic expansions are useful. By asymptotic expansions we mean the solutions for a given size of the scale factor,  where one or more of the constituents contribute only marginally to the result. 

In the following we use a looser definition of $\Omega_i$, where at each size of the scale factor $\Omega_i$ denotes the fractional contribution to $H(a)$. That is, for matter for example we re-define $\Omega_m(a,r)=\frac{8 \pi \Mtsq}{3 H(a)^2} \frac{1}{a^3}$ and so on.

We focus only on solutions with a Big Bang, and only solutions with a non-zero matter (dust) content. Then, given Eqs.~\eqref{eq:LTB}~and~\eqref{eq:t_LTB_full}, which we repeat here,
\begin{align}
\left( \frac{\dot R}{R} \right)^2 =  \left( \frac{\dot a}{a} \right)^2 = & H^2(r,t)=\frac{8 \pi \Mtsq}{3}\left[   \frac{1}{a^3(r,t)} +  \frac{3\kappa(r)}{4\pi a^2(r,t)}  + \frac{\Lambda}{8\pi\Mtsq}    \right],\\
t(A) - t_{BB}(r) =&\frac{1}{\sqrtMtsq}\int_0^{A}\frac{\sqrt{\tilde a}\,d\tilde a}{\sqrt{ \frac{8 \pi}{3} +  2 \kappa(r) {\tilde a}  + \frac{\Lambda}{3\Mtsq}  {\tilde a^3 }}},\label{eq:t_LTB_full_app}
\end{align}
we see that for any choice of $\kappa(r)$ and $\Lambda$, prior to some initial time the equation is dominated by matter. Hence the integral that gives $t(a)$ always has a non-negligible contribution from the matter content, even for $a$ so large that $\frac{1}{a^3}\ll  \frac{\Lambda}{8\pi\Mtsq}$. Therefore we do not consider asymptotic expansions for $\Omega_m\ll 1$.

Similarly, when expanding the integrand in Eq.~\eqref{eq:t_LTB_full_app} for small omega's, it only makes sense to expand in $\Omega_k$ if $\Omega_k\ll \Omega_m$, regardless of $\Omega_\Lambda$. To summarize, one can use expansions under the conditions,
\begin{align}
\mbox{expand in $\Omega_k$ when}&&\frac{3 \kappa(r) a}{4\pi} <& \epsilon,\\
\mbox{expand in $\Omega_\Lambda$ when}&&\frac{\Lambda a^3}{8\pi \Mtsq + 6 \kappa(r) a} <& \epsilon,
\end{align}
guaranteeing that the expansion is valid throughout the whole integration from $0$ to $a$, for $t(a)$ . When one uses a linear expansion, one should set $\epsilon$ to $\epsilon = \sqrt{\eta}$, with $\eta$ the desired accuracy, such that the error is $\mathcal{O}(\epsilon^2)=\mathcal{O}(\eta)$. In the case of numerical computation, one sets $\eta$ to the machine precision, which for double precision (64 bit floating point) is $\eta=10^{-16}$, such that $\epsilon=10^{-8}$. That is, in double precision one has approximately 16 significant digits.

\subsection{$\Omega_\Lambda\ll1$, $\Omega_m\neq0\neq\Omega_k$}
For small $\Lambda$, the integral in Eq.~\eqref{eq:t_LTB_full_app} becomes
\begin{align}
\sqrtMtsq [ t(A) &- t_{BB}(r)  ] = \int_0^{A}\frac{\sqrt{\tilde a}\,d\tilde a}{\sqrt{ \frac{8 \pi}{3} +  2 \kappa(r) {\tilde a} }}-
\frac{\Lambda}{6\Mtsq}\int_0^{A}\frac{{\tilde a^{\frac{7}{2}} }\,d\tilde a}{\left( \frac{8 \pi}{3} +  2 \kappa(r) {\tilde a}  \right)^{\frac{3}{2}}}
+\mathcal{O}\left(\frac{\Lambda^2}{\Mtsqsq}\right)\nonumber\\
=&  \kappa(r)^{-\frac{3}{2}}  \left\{ \sqrt{\frac{A \kappa(r)}{6}}\sqrt{3 A \kappa(r) + 4\pi}-\frac{2^{\frac{3}{2}}\pi}{3}\sinh ^{-1}\left(\sqrt{\frac{3 A \kappa(r)}{4\pi}}\right)\right\}\nonumber\\
&  -\frac{\Lambda}{6\Mtsq}\kappa(r)^{-\frac{9}{2}}\left\{\frac{\sqrt{6} \sqrt{A \kappa(r)} \left(9 A^3 \kappa(r)^3-21 \pi  A^2 \kappa(r)^2+70 \pi ^2 A \kappa(r)+280 \pi
   ^3\right)}{108 \sqrt{3 A \kappa(r)+4 \pi }} \right.\nonumber\\ 
   &\hspace{1cm} \left. -{\frac{70}{27} \pi ^3 \sqrt{2} \sinh ^{-1}\left(\sqrt{\frac{3 A \kappa(r)}{4\pi}}\right)}\right\}+\mathcal{O}\left(\frac{\Lambda^2}{\Mtsqsq}\right),
\end{align}
which reduces to the well-known result as $\Lambda\rightarrow0$. The turning point from expansion to collapse lies at $a=-\frac{4\pi}{3\kappa(r)}$ if this expression is positive. Care must be taken with the branch cut of the square root here, as for negative $\kappa$, $\left(\frac{1}{\kappa}\right)^{\frac{3}{2}} =- \kappa^{-\frac32}$. For $\kappa^{-\frac92}$ the minus signs cancel, and no care has to be taken.

To obtain the derivative of time with respect to radius $r$ in this case, one must first take the derivative of the full expression, and then expand, to arrive at,
\begin{align}
\sqrtMtsq \partial_r \left[ t(A) - t_{BB}(r) \right ] =& \frac{a'(r,t) \sqrtMtsq}{ \dot a(r,t)} \nonumber\\ 
&-\frac{\kappa'(r)}{\sqrt{2} \kappa(r)^{\frac52} }\left\{ \frac{ \sqrt{3} (A \kappa(r))^{3/2}+4 \sqrt{3} \pi  \sqrt{A \kappa(r)}}{4  \sqrt{3 A \kappa(r)+4 \pi }} \right.\nonumber\\
&\left.\hspace{1cm} -\pi \sinh ^{-1}\left(\sqrt{\frac{3 A \kappa(r)}{4\pi}}\right)\right\}\nonumber\\
&+\frac{\Lambda  }{\Mtsq  }  \frac{\kappa'(r)  }{\sqrt{2}\kappa(r)^{\frac72} }\left\{ \frac{  \sqrt{3 A \kappa(r)} \left(9 A^2 \kappa(r)^2+80 \pi  A \kappa(r)+80
   \pi ^2\right)}{48 
    (3 A \kappa(r)+4 \pi )^{\frac32}}\right.\nonumber\\
&\left.\hspace{1cm}- \frac{5 \pi   }{12 }\sinh ^{-1}\left(\sqrt{\frac{3 A \kappa(r)}{4\pi}}\right)\right\}.
\end{align}

\subsection{$\Omega_k\ll1$, $\Omega_m\neq0\neq\Omega_\Lambda$}
For small $\kappa(r)$, the integral in Eq.~\eqref{eq:t_LTB_full_app} becomes
\begin{align}
\sqrtMtsq \left[ t(A) - t_{BB}(r) \right ] =& \int_0^{A}\frac{\sqrt{\tilde a}\,d\tilde a}{\sqrt{ \frac{8 \pi}{3}   + \frac{\Lambda}{3\Mtsq}  \tilde a^3 }}-
 \kappa(r) \int_0^{A}\left[\frac{{\tilde a }}{ \frac{8 \pi}{3} +   \frac{\Lambda}{3\Mtsq}  \tilde a^3}   \right]^{\frac{3}{2}}\,d\tilde a
+\mathcal{O}\left(\kappa(r)^2\right).
\end{align}
The first integral is in principle elliptic, and the full solution in Eq.~\eqref{eq:t_LTB_full_soln} is applicable when one defines the correct roots $z_i$, however this special case is has a known solution in terms of $\sinh(x)$ and can be found in the literature. The second integral is of course the integral that is solved in the main text for $t'(a,r)$ for the special case where $\kappa(r)=0$. Hence,
\begin{align}
\sqrtMtsq &\left[ t(A) - t_{BB}(r) \right ] = {\sqrt{\frac{4 \sqrtMtsq ^2}{3\Lambda }}\sinh ^{-1}\left(\frac{A^{3/2}}{2 \sqrt{2 \pi }} \sqrt{\frac{\Lambda
   }{\Mtsq}}\right)}\nonumber\\
& +\frac{2 \kappa(r)}{3}\left(\frac{\Lambda}{3\Mtsq}\right)^{\frac{3}{2}}\frac{i}{( x_1x_2x_3)^{\frac{1}{2}}}\left[\frac{1}{(x_1-x_2)(x_1-x_3)}R_D\left(\frac{1}{A}-\frac{1}{x_2}, \frac{1}{A}-\frac{1}{x_3},\frac{1}{A}-\frac{1}{x_1}\right) \right. \nonumber\\
&+\frac{1}{(x_2-x_1)(x_2-x_3)}R_D\left(\frac{1}{A}-\frac{1}{x_3}, \frac{1}{A}-\frac{1}{x_1},\frac{1}{A}-\frac{1}{x_2}\right)  \nonumber\\
&\left.+\frac{1}{(x_3-x_1)(x_3-x_2)}R_D\left(\frac{1}{A}-\frac{1}{x_1}, \frac{1}{A}-\frac{1}{x_2},\frac{1}{A}-\frac{1}{x_3}\right) \right],
\end{align}
where $x_i$ are the three solutions of $ \frac{8 \pi}{3}   + \frac{\Lambda}{3\Mtsq}  \tilde x_i^3 =0$. If one allows for a negative $\Lambda$, also this scenario can have a postive real $x_i$ at which the transition from expansion to collapse occurs. Of course care has to be taken at the limit of $\dot a \rightarrow 0$, identical to what is described in the main text concerning $a'(r,t)$.

Unfortunately, an expansion to obtain $\partial_r \left[ t(A) - t_{BB}(r) \right ]$ in this asymptotic region is not trivial, as,
\begin{align}
\sqrtMtsq &\partial_r \left[ t(A) - t_{BB}(r) \right ] = \frac{a'(r,t) \sqrtMtsq}{ \dot a(r,t)} \nonumber\\ 
&+\frac{2 \kappa'(r)}{3}\left(\frac{\Lambda}{3\Mtsq}\right)^{\frac{3}{2}}\frac{i}{( x_1x_2x_3)^{\frac{1}{2}}}\left[\frac{1}{(x_1-x_2)(x_1-x_3)}R_D\left(\frac{1}{A}-\frac{1}{x_2}, \frac{1}{A}-\frac{1}{x_3},\frac{1}{A}-\frac{1}{x_1}\right) \right. \nonumber\\
&+\frac{1}{(x_2-x_1)(x_2-x_3)}R_D\left(\frac{1}{A}-\frac{1}{x_3}, \frac{1}{A}-\frac{1}{x_1},\frac{1}{A}-\frac{1}{x_2}\right)  \nonumber\\
&\left.+\frac{1}{(x_3-x_1)(x_3-x_2)}R_D\left(\frac{1}{A}-\frac{1}{x_1}, \frac{1}{A}-\frac{1}{x_2},\frac{1}{A}-\frac{1}{x_3}\right) \right]\nonumber\\
& + {3 \kappa(r) \kappa'(r)} \int_0^{A}\left[\frac{{\tilde a }}{ \frac{8 \pi}{3} +   \frac{\Lambda}{3\Mtsq}  \tilde a^3}   \right]^{\frac{5}{2}}\,d\tilde a
+\mathcal{O}\left(\kappa(r)^2\right),
\end{align}
where at this moment we do not know of a solution of the last integral in terms of symmetric elliptic integrals, so we leave that integral for future work.

\subsection{$\Omega_\Lambda\ll1$, $\Omega_k\ll1$, $\Omega_m\neq0$}
The simplest of the expansions is the scenario with small $\kappa(r)$ and $\Lambda$. The integral becomes,
\begin{align}
\sqrtMtsq \left[ t(A) - t_{BB}(r) \right]=&\sqrt{ \frac{3}{8 \pi} }\int_0^{A}d\tilde a \left[\sqrt{\tilde a} - \frac{3}{8 \pi} \kappa(r) \tilde a^{\frac{3}{2}} - \frac{3}{8 \pi} \frac{\Lambda}{6\Mtsq} \tilde a^{\frac{7}{2}}\right]  + \mathcal{O}\left(\kappa(r)^2,\Lambda^2,\kappa(r)\Lambda\right)\nonumber\\
=\frac{2}{3}\sqrt{ \frac{3}{8 \pi} }A^{\frac32}&-\frac25\left( \frac{3}{8 \pi} \right)^{\frac{3}{2}}\kappa(r) A^{\frac{5}{2}} -\frac29\left( \frac{3}{8 \pi} \right)^{\frac{3}{2}}\frac{\Lambda}{6\Mtsq} A^{\frac{9}{2}}+ \mathcal{O}\left(\kappa(r)^2,\Lambda^2,\kappa(r)\Lambda\right),
\end{align}
with the spatial derivative,
\begin{align}
\sqrtMtsq \partial_r [ t(A)& - t_{BB}(r) =\nonumber\\
& \frac{a'(r,t) \sqrtMtsq}{ \dot a(r,t)} + \kappa'(r) \left\{ 
-\frac{2}{5}\left( \frac{3}{8 \pi} \right)^{\frac{3}{2}}A^{\frac52}
+\frac67\left( \frac{3}{8 \pi} \right)^{\frac{5}{2}}\kappa(r) A^{\frac{7}{2}} 
+\left( \frac{3}{8 \pi} \right)^{\frac{5}{2}} \frac{\Lambda}{11 \Mtsq} A^{\frac{11}{2}} \right\}.
\end{align}

\section{Solution in presence of radiation\label{app:radsoln}}
Writing $A\equiv a(r,t)$, the general solution to $t$ for the Friedmann equations in presence of radiation, matter, curvature and a cosmological constant is given by the integral,
\begin{align}
\int_0^{A}&\frac{\tilde a\,d\tilde a}{\sqrt{ \Omega_r + {\Omega_m}{\tilde a} +  {\Omega_k}{\tilde a^2}  + {\Omega_\Lambda} {\tilde a^4 }}}=H_*\left[t(A) - t_{BB}(r)\right],
\end{align}
which is the same as Eq.~\eqref{eq:tofa_prod_0_A} for $n=4$.

We split the integral in two parts, where it is most convenient to take $z_1$ to be the smallest positive real root, if any root is positive and real (otherwise any root will do),
\begin{align}
H_* \left[t(A) - t_{BB}\right]=&\frac{1}{\sqrt{\Omega_\Lambda}}\int_0^{A} \frac{(a -z_1 + z_1) \,d\,a}{\sqrt{\prod_{m=1}^4 (a-z_m)}}\nonumber\\
=&\frac{1}{\sqrt{\Omega_\Lambda}}\int_0^{A} \frac{(a -z_1 )\, d\,a}{\sqrt{\prod_{m=1}^4 (a-z_m)}}+\frac{z_1}{\sqrt{\Omega_\Lambda}}\int_0^{A} \frac{  d\,a}{\sqrt{\prod_{m=1}^4 (a-z_m)}}\label{eq:n4_full}
\end{align}
In the second integral we can substitute $a\rightarrow b=\frac{1}{a}-\frac{1}{A}$, 
\begin{align}
\frac{z_1}{\sqrt{\Omega_\Lambda}}\int_0^{A} \frac{  d\,a}{\sqrt{\prod_{m=1}^4 (a-z_m)}}=&\frac{1}{\sqrt{\Omega_\Lambda}}\frac{\sqrt{z_1}}{\sqrt{\prod_{m=2}^4 z_m}} \int_{0}^{\infty} \frac{d\,b}{\sqrt{\prod_{m=1}^4 (b+\frac{1}{A}-\frac{1}{z_m})}}\nonumber\\
=&\frac{1}{\sqrt{\Omega_\Lambda}}\frac{\sqrt{z_1}}{\sqrt{\prod_{m=2}^4 z_m}} R_F(V^2_{12},V^2_{13},V^2_{23}).\label{eq:n4_simple}
\end{align}
where $V_{ij}=\sqrt{\frac{1}{A}-\frac{1}{z_i}}\sqrt{\frac{1}{A}-\frac{1}{z_j}}-\sqrt{\frac{1}{A}-\frac{1}{z_k}}\sqrt{\frac{1}{A}-\frac{1}{z_l}}$ with $\{i,j,k,l\}=\{1,2,3,4\}$ ({\em e.g.} if $i,j=2,3$, then $k,l=1,4$). The last equality in Eq.~\eqref{eq:n4_simple} is proven in Ref.~\cite{carlson97_symbolic}, and the equality is invariant under the ordering of the roots $z_m$, that is, invariant under the choice for $z_4$.

The first integral in Eq.~\eqref{eq:n4_full} takes another road. Following Ref.~\cite{Carlson1999739} we make the change of variables $a\rightarrow\frac{A\, b}{b+1}$,  
\begin{align}
\frac{1}{\sqrt{\Omega_\Lambda}}\int_0^{A}& \frac{(a -z_1 )\, d\,a}{\sqrt{\prod_{m=1}^4 (a-z_m)}}=\frac{1}{\sqrt{\Omega_\Lambda}}\int_0^\infty \frac{\left(\frac{A\,b}{b+1}-z_1\right) \frac{A}{(b+1)^2}\,db}{\prod_{m=1}^4 \sqrt{\frac{A\, b}{b+1}-z_m}}\nonumber\\
=& \frac{1}{\sqrt{\Omega_\Lambda}}\int_0^\infty \frac{\left((A-z_1)b - z_1\right)\,db}{(b+1)\prod_{m=1}^4 \sqrt{(A-z_m)b-z_m}}\nonumber\\
=& \frac{1}{\sqrt{\Omega_\Lambda}}\sqrt{\frac{A-z_1}{\prod_{m=2}^4 (A-z_m)}}\int_0^\infty \frac{\left(b - \frac{z_1}{A-z_1}\right)\,db}{(b+1)\prod_{m=1}^4 \sqrt{b-\frac{z_m}{A-z_m}}}\nonumber\\
=& \frac{1}{\sqrt{\Omega_\Lambda}}\sqrt{\frac{A-z_1}{\prod_{m=2}^4 (A-z_m)}} \left[   \frac{2z_{12}z_{13}z_{14}}{3 (z_1-1)}  R_J\left( W_{12}^2, W_{13}^2,W_{23}^2,W_{10}^2\right) +2 R_C\left( \alpha^2,\beta^2     \right)\right].\label{eq:n4_complic}
\end{align}
In the last line we use the definitions,
\begin{align}
W_{ij}=&\sqrt{- \frac{z_i}{A-z_i}}\sqrt{- \frac{z_j}{A-z_j}}+\sqrt{- \frac{z_k}{A-z_k}}\sqrt{- \frac{z_l}{A-z_l}}\hspace{.5cm}\mbox{with }\{i,j,k,l\}=\{1,2,3,4\},\\
z_{ij}=&z_i-z_j\\
W_{10}^2=&W_{ij}^2-\frac{z_{ik}z_{il}(z_j-1)}{z_1-1},\\
\alpha=&1+\frac{\sqrt{z_2}\sqrt{ z_3}\sqrt{ z_4}}{\sqrt{z_1}},\\
\beta^2=&\frac{1}{z_1}W_{10}^2.
\end{align}

In Eq.~\eqref{eq:n4_complic} we wrote the intermediate step in the second line explicitly, since from there it is straightforward to see that for real positive $z_1$, we have
\begin{align}
\lim_{A\rightarrow z_1}\frac{1}{\sqrt{\Omega_\Lambda}}\int_0^{A} \frac{(a -z_1 )\, d\,a}{\sqrt{\prod_{m=1}^4 (a-z_m)}}=&\frac{2\sqrt{z_1}}{3\sqrt{\Omega_\Lambda}}\sqrt{\frac{1}{\prod_{m=2}^4 (z_1-z_m)}}R_J\left( -\frac{z_2}{z_1-z_2},-\frac{z_3}{z_1-z_3},-\frac{z_4}{z_1-z_4},1  \right).
\end{align}
The second integral in Eq.~\eqref{eq:n4_full} is continuous in this limit.

Alltogether this gives the final result,
\begin{align}
H_* \left[t(A) - t_{BB}\right]=& \frac{1}{\sqrt{\Omega_\Lambda}}\sqrt{\frac{A-z_1}{\prod_{m=2}^4 (A-z_m)}} \left[   \frac{2z_{12}z_{13}z_{14}}{3 (z_1-1)}  R_J\left( W_{12}^2, W_{13}^2,W_{23}^2,W_{10}^2\right) +2 R_C\left( \alpha^2,\beta^2     \right)\right]\nonumber\\ &+\frac{1}{\sqrt{\Omega_\Lambda}}\frac{\sqrt{z_1}}{\sqrt{\prod_{m=2}^4 z_m}} R_F(V^2_{12},V^2_{13},V^2_{23}).
\end{align}

\subsection{Roots $z_i$}
In order to write down the roots $\{z_i\}$, let us use the following definitions,
\begin{align}
X=&\frac{\Omega_r}{\Omega_\Lambda}, \hspace{1cm} Y = \frac{{\Omega_m}}{\Omega_\Lambda}, \hspace{1cm}  Z=  \frac{{\Omega_k}}{\Omega_\Lambda},\\
K=&\left({\sqrt{\left(-72 X Z+27 Y^2+2 Z^3\right)^2-4 \left(12
   X+Z^2\right)^3}-72 X Z+27 Y^2+2 Z^3}\right)^{\frac13},\\
L=&-\frac{2Z}{3}+\frac{2^{\frac13}(12 X + Z^2)}{3K}  + \frac{K}{2^{\frac13}3}.
\end{align}
The roots are
\begin{align}
z_1=&\frac{1}{2}\sqrt{L}  -  \frac{1}{2}\sqrt{-\frac{6Z}{3} - L -\frac{2Y}{\sqrt{L}}},\\
z_2=&\frac{1}{2}\sqrt{L}  +  \frac{1}{2}\sqrt{-\frac{6Z}{3} - L -\frac{2Y}{\sqrt{L}}},\\
z_3=&\frac{1}{2}\sqrt{L}  -  \frac{1}{2}\sqrt{-\frac{6Z}{3} - L + \frac{2Y}{\sqrt{L}}},\\
z_4=&\frac{1}{2}\sqrt{L}  +  \frac{1}{2}\sqrt{-\frac{6Z}{3} - L + \frac{2Y}{\sqrt{L}}}.
\end{align}

\section{The interpolating function $W_n(x,a)$}\label{app:wfunc}
We define the function $W_n(x,a)$ as,
\begin{equation}
\begin{array}{ll}
\begin{array}{rl}
\phantom{_{+1}}W_n(x,\alpha)\equiv&1
\end{array}&\mbox{for $x<\alpha$}\\[\mystretch]
\left.
\begin{array}{rl}
W_2(x,\alpha)\equiv&\frac{1}{2}+\frac{1}{2}\sin\left[ \pi\left(\frac{1}{2}-\frac{x-\alpha}{1-\alpha}\right)\right]\\[\mystretch]
W_{n+1}(x,\alpha)\equiv& \frac{\int_0^{\frac{1-x}{1-\alpha}}W_n(\left|1-2x'\right|,0)dx'}{\int_0^{1}W_n(\left|1-2x''\right|,0)dx''}\hspace{1cm}\mbox{for $n\geq 2$}.
\end{array}
\right\}&\mbox{for $\alpha<x<1$}\\[2\mystretch]
\begin{array}{rl}
\phantom{_{+1}}W_n(x,\alpha)\equiv0
\end{array}&\mbox{for $1<x$},
\end{array}\label{eq:wfunc}
\end{equation}
which is $C^n$ everywhere. Note the absolute value inside the integral, which takes care of mirroring two images of the function that interpolates from one to zero, in order to have a function that goes from zero to one to zero. Also, inside the integral the functions are evaluated for $\alpha=0$, because the integration limits are always $0<\{x',x''\}<1$.

One can construct a similar function taking polynomials in stead of sinusoids, by starting with $W_0(x,\alpha)=(1-x)/(1-\alpha)$ for $\alpha < x < 1$.

We constructed this function by starting with the function $f(x)=1+\sin(x)$, which equals zero and has a zero derivate at $x=-\pi/2$ and at $x=3\pi/2$. Integrating this function over $x$ then leads to a function which has zero first and second derivatives at $x=-\pi/2$ and at $x=3\pi/2$. Next one matches this function too a mirror image, normalizes it to zero at both end points of the domain and integrates again, such that one ends up with a function that  has zero first, second and third derivatives at the domain borders. One can continue this scheme forever, as explicated in Eq.~\eqref{eq:wfunc}.

\section{Usage of the numerical module {\sc ColLambda}}\label{app:code}
The module {\sc ColLambda} is written in {\sc Fortran}90, and can be downloaded from \\ {\myurl}. Compilation instructions can be found at that website. The module can be included by the statement,
\begin{verbatim}
use LLTB
\end{verbatim}
and any metric function and a number of derivatives is then obtained by the function call,
\begin{verbatim}
call lltb_functions(H0_inf, Lambda, kofr, dkdr, tbbofr, dtbbdr, Mtilde, &
     r, t, &
     Rltb, Rp, Rpd, Rpdd, S, Sd, Sdd, a, ap, apd, add, apdd, H, Hp, tturn)
\end{verbatim}
where all items in the first and second line are mandatory and are {\texttt{intent(in)}}, and all items in the third line are optional and {\texttt{intent(out)}}. The normalization parameters and their corresponding parameters in this paper are
\begin{align}
{\texttt{H0\_inf}}&\,\,\,\,=H_*\equiv H^2(r_*,t_0)\nonumber\\
{\texttt{Lambda}}&\,\,\,\,=\Lambda\nonumber\\
{\texttt{Mtilde}}&\,\,\,\,=\Mtsq\nonumber
\end{align}
and the functions
\begin{align}
{\texttt{kofr}}&\,\,\,\,=\kappa(r)\nonumber\\
{\texttt{dkdr}}&\,\,\,\,=\kappa'(r)\nonumber\\
{\texttt{tbbofr}}&\,\,\,\,=t_{BB}(r)\nonumber\\
{\texttt{dtbbdr}}&\,\,\,\,=t_{BB}'(r)\nonumber
\end{align}
must each be a pure function of $r$, declared as:
\begin{verbatim}
    function kofr(r)
      real(8) :: kofr
      real(8), intent(in) :: r
    end function kofr
\end{verbatim}
As a consequence, any other parameters on which $k(r)$ may depend, such as the maximum size $L$, maximum curvature $\kappa_{\rm max}$, or anything else, must be global variables which the function {\texttt{kofr(r)}} can access without receiving them as arguments.
The values that the subroutine returns have the following notation:
\begin{align}
{\texttt{Rltb}}&\,\,\,\,=R(r,t)\nonumber&
{\texttt{Rp}}&\,\,\,\,=R'(r,t)\nonumber\\
{\texttt{Rpd}}&\,\,\,\,=\dot R'(r,t)\nonumber&
{\texttt{Rpdd}}&\,\,\,\,=\ddot R'(r,t)\nonumber\\
{\texttt{S}}&\,\,\,\,=S(r,t)\nonumber&
{\texttt{Sd}}&\,\,\,\,=\dot S(r,t)\nonumber\\
{\texttt{Sdd}}&\,\,\,\,=\ddot S(r,t)\nonumber&
{\texttt{a}}&\,\,\,\,=a(r,t)\nonumber\\
{\texttt{ap}}&\,\,\,\,=a'(r,t)\nonumber&
{\texttt{apd}}&\,\,\,\,=\dot a'(r,t)\nonumber\\
{\texttt{add}}&\,\,\,\,=\ddot a(r,t)\nonumber&
{\texttt{apdd}}&\,\,\,\,=\ddot a'(r,t)\nonumber\\
{\texttt{H}}&\,\,\,\,=H(r,t)\nonumber&
{\texttt{Hp}}&\,\,\,\,=H'(r,t)\nonumber\\
{\texttt{tturn}}&\,\,\,\,=t_{U}(r)\nonumber
\end{align}
where $t_U(r)$ denotes the time at which $\dot a(r,t_U)=0$, if it exists. If it exists, the local singularity is reached at $t=t_{BB}(r) + 2t_U$, that is, $a(r,t_{BB}(r))=a(r,t_{BB}(r)+2t_U)=0$. If this time does not exist, {\em i.e.} when the solution is ever expanding, {\texttt {tturn}} will be set to $10^{30}$.

An example call, where the user has set the normalization variables and has defined the necessary functions $\kappa(r)$, $t_{BB}(r)$ and their first derivatives, to set {\texttt{myRltb}} to $R(r,t)$, {\texttt{myS}} to $S(r,t)$ and {\texttt{myt}} to $t_U(r)$, would look like:
\begin{verbatim}
call lltb_functions(H0_inf, Lambda, kofr, dkdr, tbbofr, dtbbdr, Mtilde, &
     r, t, &
     Rltb=myRltb, S=myS,tturn=myt)
\end{verbatim}

\renewcommand{\bibfont}{\footnotesize}\bibliographystyle{unsrtnat_wv}\bibliography{refs}

\end{document}